\documentclass[a4paper,11pt]{revtex4}
\usepackage{amssymb}
\usepackage{amsmath}
\usepackage{graphics}
\usepackage{color}
\usepackage[T1]{fontenc}
\usepackage{times,mathptm}
\usepackage[cp1250]{inputenc}
\usepackage{epsfig,epsf}
\newcommand{\be}{\begin{equation}}
\newcommand{\ee}{\end{equation}}

\newcommand\beq{\begin{eqnarray}}
\newcommand\eeq{\end{eqnarray}}

\begin{document}

\title{Twist decomposition of proton structure from BFKL and BK amplitudes}


\author{Leszek Motyka} \email{leszek.motyka@uj.edu.pl}

\author{Mariusz Sadzikowski} \email{sadzikowski@th.if.uj.edu.pl}
\affiliation{M.\ Smoluchowski Institute of Physics,  Jagiellonian University,
PL~30--059 Cracow, Poland}

\date{\today}

\begin{abstract}
An analysis of twist composition of Balitsky-Kovchegov (BK) amplitude is performed in the double logarithmic limit. In this limit the BK evolution of color dipole -- proton scattering is equivalent to BFKL evolution which follows from vanishing of the Bartels vertex in the collinear limit. We perform twist decomposition of the BFKL/BK amplitude for proton structure functions and find compact analytic expressions
that provide accurate approximations for higher twist amplitudes. The BFKL/BK higher twist amplitudes are much smaller than those following from eikonal saturation models.
\end{abstract}

\maketitle

\setcounter{footnote}{0}

\vskip0.2cm

\section{Introduction}

The standard description of proton hard interactions is based on the leading twist-2 term in the Operator Product Expansion (OPE). At twist-2 the hard factorization theorem holds true for sufficiently inclusive cross-sections and the non-perturbative features of the proton structure may be absorbed into universal quantities --- parton distribution functions (pdfs). The accuracy of this powerful description is, however, limited due to possible contribution of higher twist terms in OPE. Although the higher twist terms are power suppressed by the process hard scale, they may provide sizeable corrections at moderate scales, in particular at small values of parton~$x$, where the QCD evolution leads to relative enhancement of higher twist contribution. Hence the higher twist effects may affect the fits of parton distribution functions and they contribute to theoretical uncertainty of the fitted pdfs, and it is important to estimate their magnitude both with theoretical and experimental methods. In particular, recent HERA data provided evidence for breakdown of twist-two DGLAP fits in diffractive DIS (DDIS) \cite{ZEUSddis,MSS}, which may be explained with emergence of strong, higher twist effects at $Q^2 < 5$~GeV$^2$ and at small~$x$~\cite{MSS}. Also the inclusive DIS data show significant deviations from DGLAP fits in similar kinematic region \cite{dis14talk}. Since these are the first signals of higher twist effects in proton structure functions, theoretical estimates must rely on models that would allow to constrain unknown non-perturbative matrix elements of higher twist operators.

Currently the most common scheme to estimate higher twist contributions at small~$x$ proton scattering is based on the Glauber-Mueller (eikonal) picture of multiple independent scatterings of color dipoles off proton. This simplest assumption on scatterings was employed in the very successful and efficient Golec-Biernat-W\"{u}sthoff (GBW) saturation model \cite{GBW}.  It leads to a unified description of DIS down to photoproduction limit together with diffractive DIS, so it should carry some information about higher twist terms that are important at low scales. The model was used for twist analyses of DIS, DDIS and forward Drell-Yan cross sections
\cite{twist1,twist2,twist3, MSS}.

The simple GBW assumption of multiple independent colour dipole scatterings, however, does not hold in QCD at leading logarithmic approximation. This follows, in particular, from the analysis of multiple gluon exchanges in the leading logarithmic $1/x$-limit (LL$1/x$), that is in LL BFKL formalism \cite{bfkl0, bfklrev}. At the LL$1/x$ accuracy the gluon reggeization mechanism (Regge bootstrap) reduces multiple elementary gluon exchanges with a projectile color dipole to an exchange of two Reggeized gluons, that span the BFKL ladder \cite{EGGLA}. At the double leading logarithmic limit (DLA) the BFKL and DGLAP evolutions are equivalent, so at DLA the amplitude of colour dipole scattering is driven by a single DGLAP ladder, corresponding to a single scattering and not multiple, eikonal scatterings. Using the language of anomalous dimensions, one concludes that the color dipole does not couple at the leading logarithmic approximation to states with anomalous dimensions of multiple independent DGLAP ladders at the large $N_c$~limit. It should be remembered though that the Reggeized gluon is a composite of infinitely many elementary gluons so in BFKL the color dipole still undergoes multiple scattering but the multiple exchanges are described by different anomalous dimensions than the leading anomalous dimension of multiple DGLAP ladder exchanges. Thus, the BFKL formalism provides an alternative QCD description to the GBW model of multiple gluon exchange --- and so of higher twist effects.

It is well known that at low scales and small~$x$ the BFKL amplitudes may lead to violations of unitarity and gluon recombination effects should be taken into accout.
This phenonemon was treated in the LL$1/x$ approximation for an asymmetric configuration of hard projectile (e.g.\ a virtual $\gamma^*$) and extended target (e.g.\ a nucleus) by Bartels~\cite{Bartels3p}, Balitsky~\cite{Balitsky} and Kovchegov~\cite{Kov}. This approach resulted with the Balitsky-Kovchegov (BK) evolution equation for QCD amplitudes, valid in the large $N_c$ limit. In a diagramatic representation the BK equation resums BFKL ladder fan diagrams in which mergings of two BFKL ladders occur (via a triple ladder vertex obtained by Bartels\cite{Bartels3p}), when the multi-ladder states evolve from the soft target to the hard projectile.

A natural question to address in a twist analysis of small-$x$ amplitudes is the twist content of BK amplitudes. One might expect some mixing to occur of higher twist operators with lower twist operators due to merging vertex of BFKL ladders. It was found, however, that this vertex vanishes in the collinear limit \cite{BarKut}. This means that in the LL$1/x$ approximation the triple-ladder vertex does not affect the leading order $Q^2$-evolution of QCD amplitudes. In fact, this finding agrees with a classical result of \cite{BoFKL} which shows that parton ladder merging vertices vanish in LL$Q^2$ evolution of multiple parton densities described by quasipartonic operators.  Vanishing of the merging vertex in the collinear limit combined with the coupling of the color dipole to a single BFKL ladder implies that the BK evolution of color dipole scattering off proton is described by a single BFKL ladder in the collinear limit. It does not mean that the BK unitarity corrections vanish completely --- in the $Q^2$ evolution they do modify the initial conditions, but they do not affect the evolution. Hence, the conclusion of this analysis is that in double logarithmic approximation the twist structure of BK evolution is given by the twist structure of BFKL evolution in which the unitarity correcions are included in the input of $Q^2$~evolution.

To sum up, in this paper we analyze the twist content of color dipole scattering amplitude off proton in the BK approach. As described above, for the twist evolution BK amplitude reduces to  BFKL evolution with a modified input. Hence we investigate twist corrections in proton structure functions in BFKL formalism and derive conclusions for BK amplitudes.

\section{Cross-sections in the color dipole model}

In the kinematical region of low $x$ the total cross section of virtual
photon-proton scattering may be described within the color dipole representation~\cite{NZ,GBW}
\be
\label{photon_proton_cross_section}
\sigma^{\gamma^\ast p}_{T,L}(x,Q^2) = \int d^2 \mathbf{r}\int_0^1 dz |\psi_{T,L}(z,r,Q^2)|^2 \sigma_{q\bar{q}}(x,\mathbf{r}),
\ee
where $T,L$ denotes transverse and longitudinal polarization of the photon.
At the lowest order in the electromagnetic coupling constant $\alpha_{em}$,
the photon light-cone wave functions squared take the form,
\cite{NZ, GBW}
\beq
\label{wave_functions}
|\psi_{L}(z,r,Q^2)|^2 &=& \frac{8N_c\alpha_{em}}{4\pi^2}\sum_f e_f^2 Q^2z^2(1-z)^2K_0^2(\epsilon r), \\\nonumber
|\psi_{T}(z,r,Q^2)|^2 &=& \frac{2N_c\alpha_{em}}{4\pi^2}\sum_f e_f^2 [z^2+(1-z)^2]\epsilon^2 K_1^2(\epsilon r),
\eeq
where for massless quarks $\epsilon = \sqrt{z(1-z)}Q$ and $r=|\mathbf{r}|$ and $K_{0,1}$ are Bessel-McDonald functions. The wave functions (\ref{wave_functions}) describe probability amplitude that the virtual photon of polarization $L,T$ fluctuates into a quark-antiquark pair of the transverse size $r$ and a fraction $z$ of the longitudinal light-cone photon momentum carried by the quark.
The dipole-proton cross section can be written as \cite{Kov}
\be
\label{dipole_cross_section}
\sigma_{q\bar{q}}(x,\mathbf{r}) = 2\int d^2\mathbf{b}\; N(x,\mathbf{r},\mathbf{b}) \equiv 2\pi R^2_p N(Y,\mathbf{r}),
\ee
where $N(x,\mathbf{r},\mathbf{b})$ is an imaginary part of the forward dipole - nucleon scattering amplitude and
$Y=\ln(x_{in}/x)$ is a rapidity variable counted with respect to some initial value $x_{in}$.
The last equality in (\ref{dipole_cross_section}) introduces effective radius of the proton $R_p$, which emerges after integration over impact parameter $\mathbf{b}$. The resulting function $N(x,\mathbf{r})$ fulfills Balitsky-Kovchegov (BK) equation without impact parameter dependence \cite{Kov, Balitsky}. The proton structure functions are related to the $\sigma^{\gamma^\ast p}$ by the formulae
\be
F_{T,L} = \frac{Q^2}{4\pi^2\alpha_{em}}\sigma^{\gamma^\ast p}_{T,L}\, ,\;\;\; F_2 = F_T+F_L\, .
\ee

In this paper we analyse the BFKL scattering amplitude, viewed as a linear version of the BK amplitude. The solution to BFKL equation in the Mellin space is well known \cite{bfklrev} and it leads to the folling BFKL amplitude of color dipole scattering:
\be
\label{mellin_transform_N}
N(x,\mathbf{r})=\frac{1}{2\pi i}\int_{C_f}ds\, r^{-2s} C(s) e^{\bar{\alpha}_s\chi(s)Y},\;\;\; \bar{\alpha}_s = \frac{N_c\alpha_s}{\pi},
\ee
where the integral is performed along the contour $C_f$ located in the fundamental strip of Mellin transformation. Function $C(s)$ depends on the initial condition for the BFKL or BK equation and $\chi(s)$ is the BFKL characteristic function,
\be
\chi(s) = 2\psi(1)-\psi(-s)-\psi(1+s),
\ee
where $\psi$ is digamma function. In this analysis we choose the exponential form of the initial condition, suggested by the Golec-Biernat-W\"usthoff (GBW) model \cite{GBW}:
\be
N(Y=0,\mathbf{r}) = 1-\exp{\left(-\frac{r^2Q_0^2}{4}\right)},
\ee
which gives $C(s) = -\Gamma(s)(4/Q_0^2)^s$ and as a possible choice of the integration contour a parallel line to imaginary axis $C_f = (-1/2-i\infty,-1/2+i\infty)$.

Twist decomposition of the cross-sections is performed using a standard Mellin technique \cite{twist1,twist2}. Substituting (\ref{mellin_transform_N})
into (\ref{photon_proton_cross_section}) one obtains
\be
\label{photon_proton_cross_section_mellin}
\sigma^{\gamma^\ast p}_{T,L}(x,Q^2) = \frac{1}{2\pi i}\int_{C_f}ds \left(\frac{Q_0^2}{Q^2}\right)^{-s} H_{T,L}(-s)\tilde{\sigma}_{q\bar{q}}(s,Y),
\ee
where the Mellin transform of photon wave functions $H_{T,L}$ can be found in \cite{twist1,twist2} and
\be
\tilde{\sigma}_{q\bar{q}}(s,Y) = -2\pi R_p^2\Gamma(s) e^{\bar{\alpha}_s\chi(s)Y}.
\ee
Mellin singularities of this amplitude give contributions to the twist expansion. They may be isolated with standard techniques of complex analysis. One closes the inverse  Mellin integration contour $C_f$ by a left semicircle at complex infinity. The integral over this closed contour is then equal to a sum of contours enevelopin the singularities, $\bigcup_{n} C_{-n}$ for $n=1,2,...$. $C_{-n}$ is a small circle of radius $\epsilon$ around negative integer $-n$. This procedure yields twist decomposition of the integral (\ref{photon_proton_cross_section_mellin})
\be
\sigma^{\gamma^\ast p}_{T,L}(x,Q^2) = \sum_{n=1}^\infty \sigma^{(2n)}_{T,L}(x,Q^2),
\ee
where twist contribution $\sigma^{(2n)}_{T,L}$ is given by the formula (\ref{photon_proton_cross_section_mellin}) with $C_f=C_{-n}$.

\section{Twist decomposition}

Expressions for twist coefficients read
\be
\label{sigma_n}
\sigma^{(2n)}_{T,L}(x,Q^2) = -R_p^2 e^{-n t} \int_{0}^{2\pi} d\theta h^{(n)}
\exp\left(\epsilon t e^{i\theta}+\frac{\bar{\alpha}_sY}{\epsilon}e^{-i\theta}\right),
\ee
where $t = \ln Q^2/Q_0^2$, function
\be
h^{(n)} = \epsilon e^{i\theta} H_{T,L}\left(n-\epsilon e^{i\theta}\right)
\Gamma\left(-n+\epsilon e^{i\theta}\right) e^{\bar{\alpha}_sY\chi_{reg}^{(n)}},
\ee
and
\be
\chi_{reg}^{(n)} = \chi\left(-n+\epsilon e^{i\theta}\right) - \frac{e^{-i\theta}}{\epsilon},
\ee
is a regular function in the limit of $\epsilon\rightarrow 0$. One can expand function $h^{(n)}$ into a series
\be
\label{hn_series}
h^{(n)} = A_0\sum_{m=-1}^\infty a_m^{(2n)T,L} \left(\epsilon e^{i\theta}\right)^m ,
\ee
where $A_0=N_c\alpha_{em}\sum_f e_f^2/\pi$ and the most singular element scales as $1/\epsilon$.
Coefficients $a_m^{(2n)T,L}$ are functions of $\bar{\alpha}_sY$ only and they are independent of $t$. Substituting (\ref{hn_series}) into equation (\ref{sigma_n}) one finds that the only integrals left are of the form
\be
\label{integral}
\int_0^{2\pi} d\theta \exp\left(\epsilon t e^{i\theta}+\frac{\bar{\alpha}_sY}{\epsilon}e^{-i\theta}+i m\theta\right) =
\left(\frac{\bar{\alpha}_sY}{\epsilon^2 t}\right)^{m/2}I_{|m|}(2\sqrt{\bar{\alpha}_sYt}),\;\;\; m=-1,0,1,2,....
\ee
where $I_m$ are modified Bessel functions of the first kind. Thus, using (\ref{hn_series}) and (\ref{integral}) one can show that
\beq
\label{sigma_series}
\sigma^{(2n)}_{T,L}(x,Q^2) &=&
-2\pi R_p^2A_0\left(\frac{Q_0}{Q}\right)^{2n}\sum_{m=-1}^\infty
a_m^{(2n)T,L}\left(\frac{\bar{\alpha}_sY}{t}\right)^{m/2}I_{|m|}(2\sqrt{{\alpha}_sYt}) \\\nonumber
&=& -2\pi R_p^2A_0\left(\frac{Q_0}{Q}\right)^{2n}\left[a_{-1}^{(2n)T,L}t\,_0\tilde{F}_1(2,\bar{\alpha}_sYt)+\sum_{m=0}^\infty
a_m^{(2n)T,L}\left(\bar{\alpha}_sY\right)^{m}\,_0\tilde{F}_1(1+m,\bar{\alpha}_sYt)\right],
\eeq
where $_0\tilde{F}_1$ is a regularized confluent hypergeometric function $_0\tilde{F}_1(m,x) =\, _0F_1(m;x)/\Gamma(m)$. The convenience of the above series follows from its fast convergence. In the experimentally interesting, broad kinematical range $2<Y<7$ and $1<t<10$ already first five terms of (\ref{sigma_series}) reproduce the exact result with accuracy better than 1\%. Hence in numerical calculations below we use first five terms, which gives sufficient accuracy for our purposes.

Performing the calculations one finds coefficients for twist~2:
\beq
\label{coefficients2}
&&a^{(2)T}_{-1} = -\frac{1}{3},\;\; a^{(2)T}_0 = -\frac{1+6\gamma_E}{18},\;\;
a^{(2)T}_1 = - \frac{112-3\pi^2+6\gamma_E(1 + 3\gamma_E)}{108}  - \frac{2}{3}\zeta(3)\bar{\alpha}_sY, \\\nonumber
&&a^{(2)T}_2 = -\frac{124- 3\pi^2 +  6\gamma_E(112 + 3\gamma_E + 6\gamma_E^2 - 3 \pi^2) +  72 \zeta(3)}{648}-\frac{1}{9}(1 + 6\gamma_E)\zeta(3) \bar{\alpha}_sY , \\\nonumber
&&a^{(2)L}_{-1} = 0,\;\; a^{(2)L}_0 = -\frac{1}{3},\;\; a^{(2)L}_1 = - \frac{-4 + 3\gamma_E}{9}, \;\;
a^{(2)L}_2 = - \frac{148 - 3\pi^2 + 6\gamma_E(3\gamma_E-8)}{108}  - \frac{2}{3}\zeta(3)\bar{\alpha}_sY .
\eeq
and coefficients for twist-4:
\beq
\label{coefficients4}
&&a^{(4)T}_{-1} = 0,\;\; a^{(4)T}_0 = -\frac{1}{5}e^{-2\bar{\alpha}_sY},\;\;
a^{(4)T}_1 = -e^{-2\bar{\alpha}_sY}\left(\frac{37 + 30\gamma_E}{150}  - \frac{2}{5}\zeta(3)\bar{\alpha}_sY\right), \\\nonumber
&&a^{(4)T}_2 = -e^{-2\bar{\alpha}_sY}\left(\frac{2144 + 30\gamma_E(37 + 15\gamma_E) - 75\pi^2}{4500}
-\frac{67 + 30\gamma_E - 30 \zeta(3)}{75} \bar{\alpha}_sY+\frac{2}{5}(\bar{\alpha}_sY)^2\right) ,\\\nonumber
&&a^{(4)L}_{-1} = \frac{4}{15}e^{-2\bar{\alpha}_sY},\;\;
a^{(4)L}_0 = e^{-2\bar{\alpha}_sY}\left(\frac{4(1 + 15\gamma_E)}{225}-\frac{8}{15}\bar{\alpha}_sY\right),\\\nonumber
&&a^{(4)L}_1 = e^{-2\bar{\alpha}_sY}\left(\frac{949 + 30\gamma_E(2 + 15\gamma_E) - 75\pi^2}{3375}
-\frac{8(16 + 15\gamma_E - 15\zeta(3))}{225}\bar{\alpha}_sY+\frac{8}{15}(\bar{\alpha}_sY)^2\right) .
\eeq
We do not display explicit expression for $a^{(4)L}_2$ coefficient because of rather lengthy formula, but is used in the numerical estimates. Let us notice $\exp (-n\alpha_sY)$ factor which is present in the higher twist-($2n$) coefficients. This term is responsible for the suppression of the higher twist terms at small $x$ values, contrary to expectations of the eikonal approach. In the double logarithmic limit
\be
_0\tilde{F}_1(1+m,\bar{\alpha}_sYt) \approx \frac{e^{2\sqrt{\bar{\alpha}_sYt}}}{2\sqrt{\pi}(\bar{\alpha}_sYt)^{(1+2m)/4}}
\ee
and using the lowest order non-zero expressions from (\ref{coefficients2}), (\ref{coefficients4}) one obtains approximate formulae for twist components:
\beq
\sigma^{(2)}_T &=& R_p^2A_0\sqrt{\pi}\left(\frac{Q_0}{Q}\right)^2\frac{t^{1/4} e^{2\sqrt{\bar{\alpha}_sYt}}}{3(\bar{\alpha}_sY)^{3/4}},\;\;\;
\sigma^{(2)}_L = R_p^2A_0\sqrt{\pi}\left(\frac{Q_0}{Q}\right)^2\frac{e^{2\sqrt{\bar{\alpha}_sYt}}}{3(\bar{\alpha}_sYt)^{1/4}}, \\\nonumber
\sigma^{(4)}_T &=& R_p^2A_0\sqrt{\pi}\left(\frac{Q_0}{Q}\right)^4\frac{e^{2\sqrt{\bar{\alpha}_sYt}-2\bar{\alpha}_sY}}{5(\bar{\alpha}_sYt)^{1/4}},\;\;\;
\sigma^{(4)}_L = -R_p^2A_0\sqrt{\pi}\left(\frac{Q_0}{Q}\right)^4\frac{4 t^{1/4} e^{2\sqrt{\bar{\alpha}_sYt}-2\bar{\alpha}_sY}}{15(\bar{\alpha}_sY)^{3/4}} .
\eeq
The above expressions coincide with the saddle point approximation of integrals $\sigma^{(2n)}_{T,L}$ \cite{lelek}.

\section{Numerical results}

With our choice of the initial condition the dipole cross-section in the BK framework depends on 4 parameters: the fixed strong coupling constant $\bar{\alpha}_s$, the initial saturation scale $Q_0$ assumed at $x=x_{in}$, and the effective proton radius $R_p$. For the present study, instead of fitting the data, we relate the BK parameters to the parameters of the GBW model \cite{GBW} assuming that dipole cross-sections coincide with each other at the initial point $x=x_{in}$
\be
2\pi R_p^2 \left(1-\exp{\left(-\frac{r^2Q_0^2}{4}\right)}\right) = \sigma_0 \left(1 - \exp{\left(-\frac{r^2Q_{GBW}(x_{in})^2}{4}\right)}\right), \\\nonumber
\ee
Thus
\beq
&&R_p^2 = \frac{\sigma_0}{2\pi},\;\;\; Q_0^2 = Q_{GBW}(x_{in})^2 = \left(\frac{x_0}{x_{in}}\right)^\lambda\; \mbox{GeV}^2,\;\;\;
\bar{\alpha}_s = \frac{\lambda}{4\ln 2} , \\\nonumber
&&\sigma_0 = 23.03\; \mbox{mb},\;\;\; \lambda = 0.288,\;\;\; x_0 = 3.04 \times 10^{-3} .
\eeq
The last equality follows from LL~BFKL relation between strong coupling constant and pomeron intercept. For numerical studies we assume $x_{in} = 0.1$.

\begin{figure}[t]
 \begin{center}
 \includegraphics[width=7.5cm]{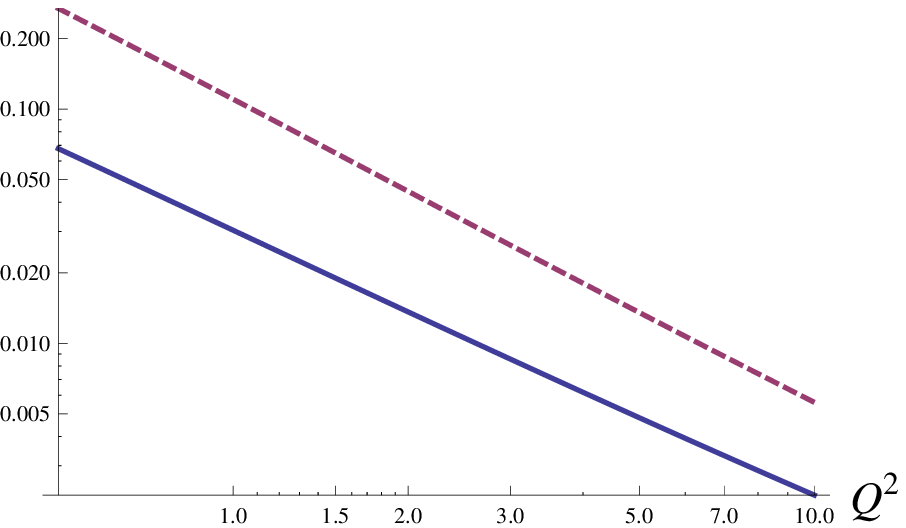}\includegraphics[width=7.5cm]{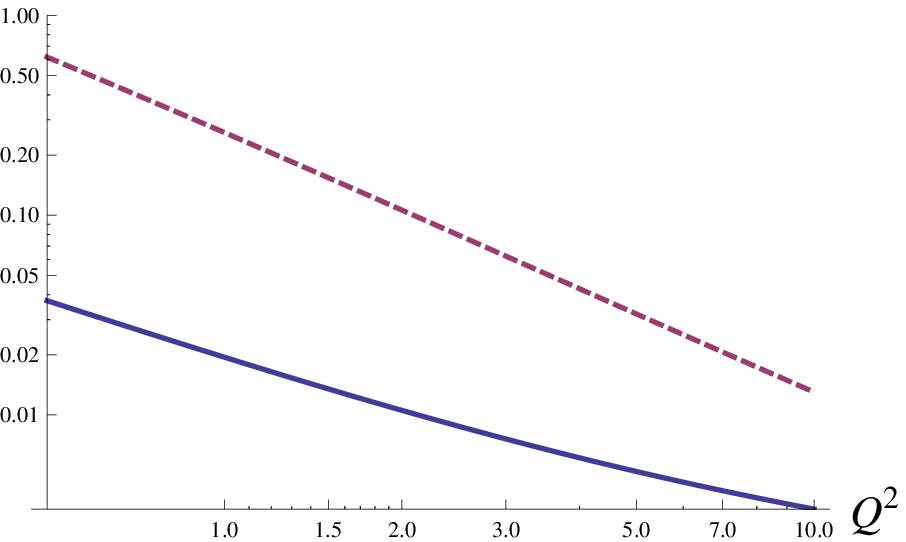}
 \caption{Relative difference $R^{(2)}_T$ between twist 2 transverse cross-section and total transverse cross-section (see text for definition) as a function
 of $Q^2$ in GeV$^2$ in log-log scale. Left panel: x=0.01. Right panel: x=0.001. The dashed curve follows from GBW model whereas the solid line from BK equation. Saturation scales from GBW model are $Q^2_{sat}=0.36, 0.71$ GeV$^2$ for $x=0.01, 0.001$ respectively.}
   \label{twist2T}
 \end{center}
\end{figure}

\begin{figure}[b]
 \begin{center}
 \includegraphics[width=7.5cm]{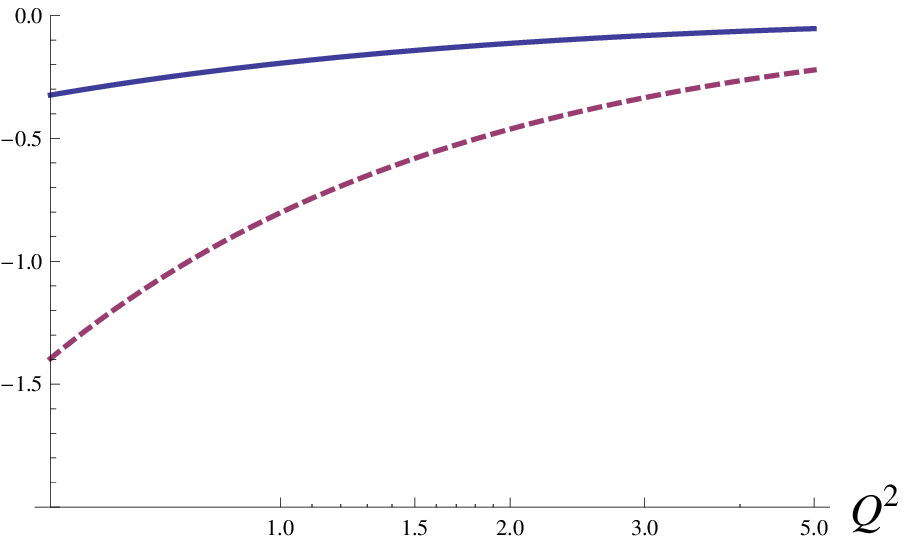}\includegraphics[width=7.5cm]{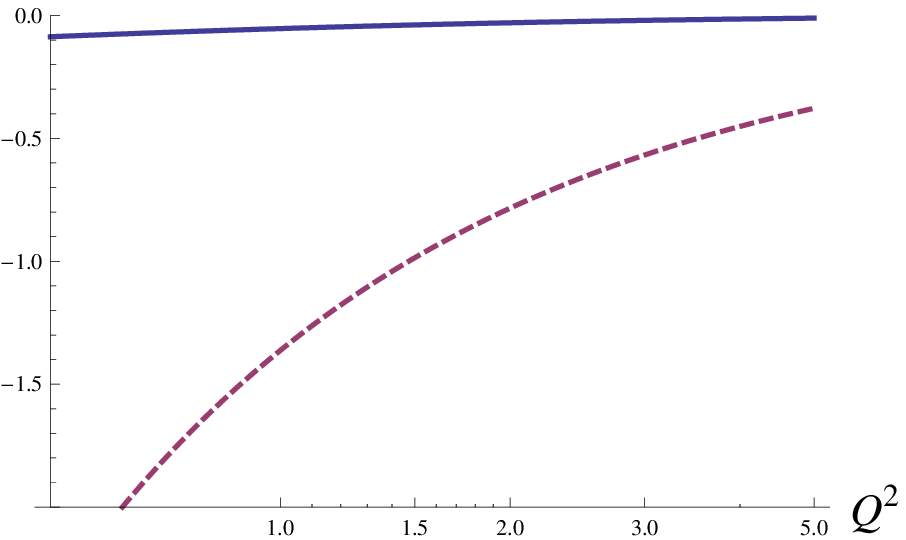}
 \caption{Relative difference $R^{(2)}_L$ between twist 2 longitudinal cross-section and total longitudinal cross-section (see text for definition) as a function
 of $Q^2$ in GeV$^2$. Left panel: x=0.01. Right panel: x=0.001. The solid line follows from BK equation whereas the dashed curve
 from GBW model. Saturation scales from GBW model are $Q^2_{sat}=0.36, 0.71$ GeV$^2$ for $x=0.01, 0.001$ respectively.}
   \label{twist2L}
 \end{center}
\end{figure}

Let us define a quantity that is a measure of a relative twist content in the total cross section,
\be
\label{R_parameter}
R^{(2,...,2k)}_{T,L} = 1-\frac{\sum_{i=1}^k\sigma^{(2i)}_{T,L}}{\sigma^{\gamma^\ast p}_{{T,L}}}
\ee
With more higher twist terms included the variable $R$ tends to zero, as the series converges to the total cross-section. In Fig.\ \ref{twist2T} we compare $R^{(2)}_T$ for GBW model and BK equation. It is clearly visible that twist-2 component of the BK cross-section makes larger part of the total cross-section than it occurs within the GBW model. This difference becomes even more striking towards smaller values of $x$, where the higher twist contribution of GBW is much larger than in the BK cross-sections. In fact, the higher twist contributions are decreasing with decreasing~$x$ in the BK approach whereas they are increasing in the GBW model. In particular for $Q^2=1$ GeV$^2$ and $x=10^{-3}$ the higher twists contribution is close to 30 per cent for the transverse structure function in GBW model and only around 2 per cent in the BK approach. Similar pattern is visible in the case of longitudinal structure function (see Fig.\ \ref{twist2L}) --- in this case the higher twist contribution dominates the result at $Q^2=1$ GeV$^2$ in the case of GBW model and it is below 10 per cent in the BK framework. Notice that the higher twist contributions are more important for the longitudinal structure function obrained from the BFKL/BK amplitudes, following the pattern found earlier within the saturation model~\cite{twist1,twist2}.

\begin{figure}[t]
 \begin{center}
 \includegraphics[width=7.5cm]{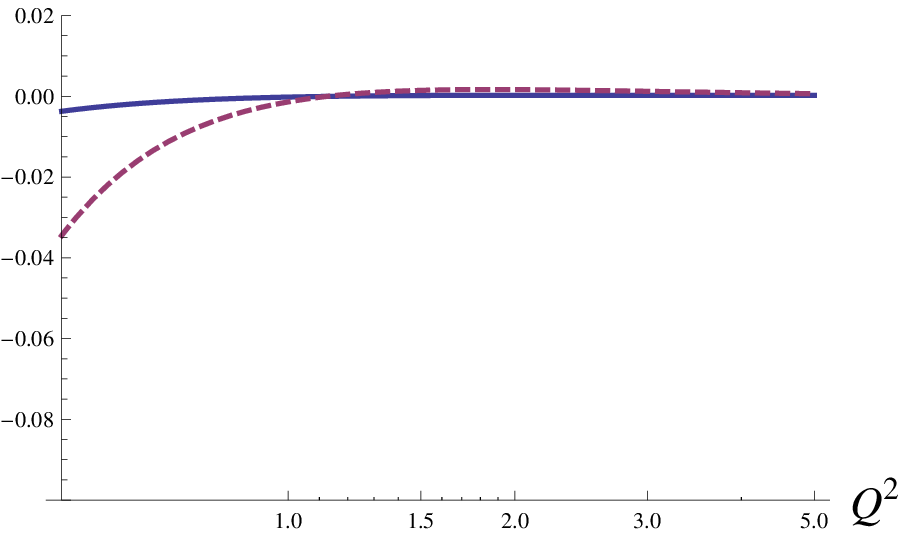}\includegraphics[width=7.5cm]{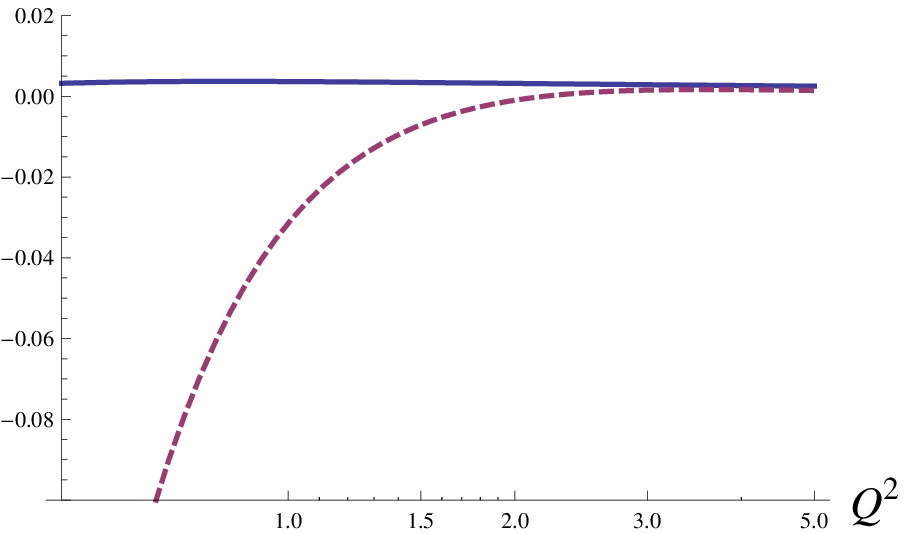}
\caption{Relative difference $R^{(2,4)}_T$ between twist 2 plus twist 4 transverse cross-section and total transverse cross-section (see text for definition)
  as a function  of $Q^2$ in GeV$^2$ in logarithmic scale. Left panel: x=0.01. Right panel: x=0.001. The solid line  follows from BK equation whereas
 the dashed curve from GBW model. Saturation scales from GBW model are $Q^2_{sat}=0.36, 0.71$ GeV$^2$ for $x=0.01, 0.001$ respectively.}
    \label{twist2and4T}
 \end{center}
\end{figure}

\begin{figure}[b]
 \begin{center}
 \includegraphics[width=7.5cm]{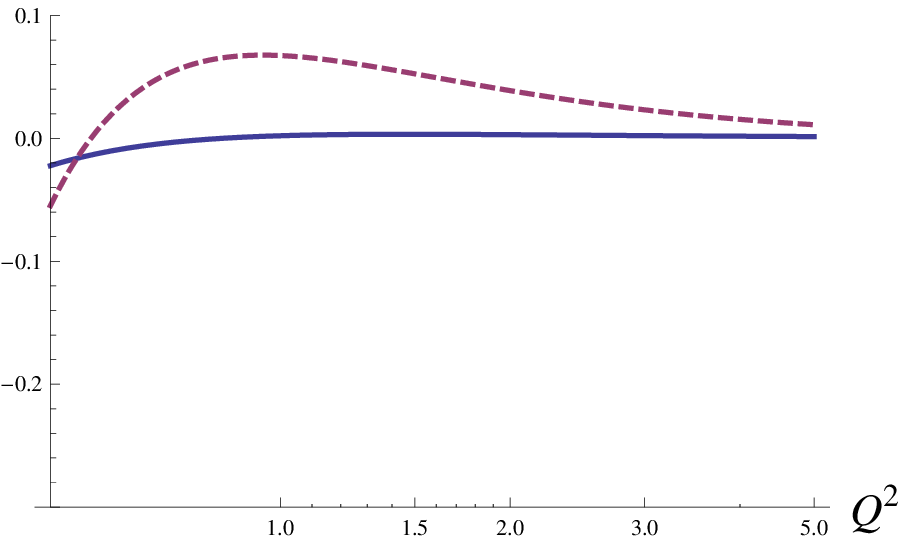}\includegraphics[width=7.5cm]{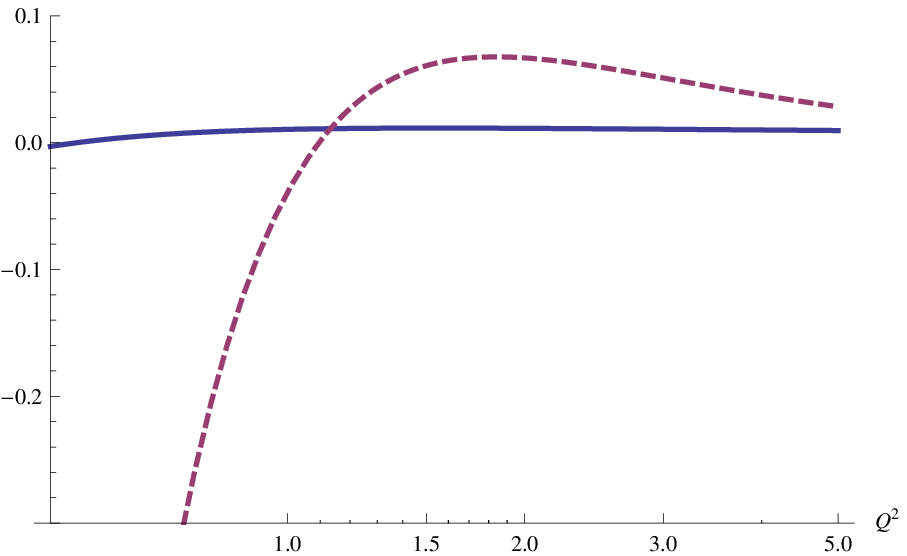}
 \caption{Relative difference $R^{(2,4)}_L$ between twist 2 plus twist 4 longitudinal cross-section and total longitudinal cross-section (see text for definition) as a function
 of $Q^2$ in GeV$^2$. Left panel: x=0.01. Right panel: x=0.001. The solid line follows from BK equation whereas the dashed curve
 from GBW model. Note different scales on vertical axis in the left and right panels.
 Saturation scales from GBW model are $Q^2_{sat}=0.36, 0.71$ GeV$^2$ for $x=0.01, 0.001$ respectively.}
   \label{twist2and4L}
 \end{center}
\end{figure}

Fig. \ref{twist2and4T} shows relative content of twist-2 and twist-4 contributions in the total transverse cross-section. For $Q^2=1$ GeV$^2$ and $x=10^{-3}$ there is only around 3~per~cent contribution left for twist-6 and higher in the case of the GBW model. This is a significant reduction of higher twist remainder compared to the 30 per cent of the higher twist remainder beyond twist 2. The contribution of twists higher than~4 is completely negligible in the BK approach. The case of the longitudinal structure function is represented in Fig.\ \ref{twist2and4L}. Similarly to the transverse case inclusion of twist-4 reduces $R$ parameter. In the GBW approach, however, the contribution from still higher twists grows quickly with decreasing $Q^2$.

\begin{figure}[t]
 \begin{center}
 \includegraphics[width=7.5cm]{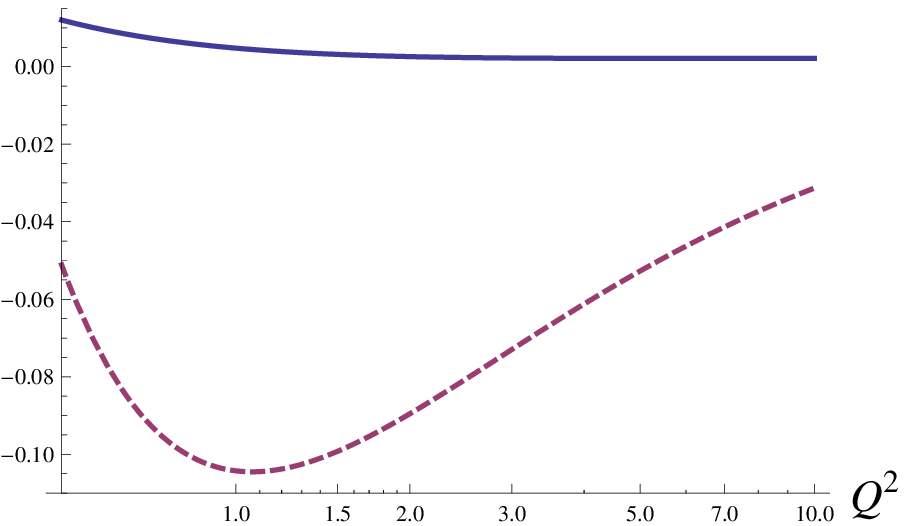}\includegraphics[width=7.5cm]{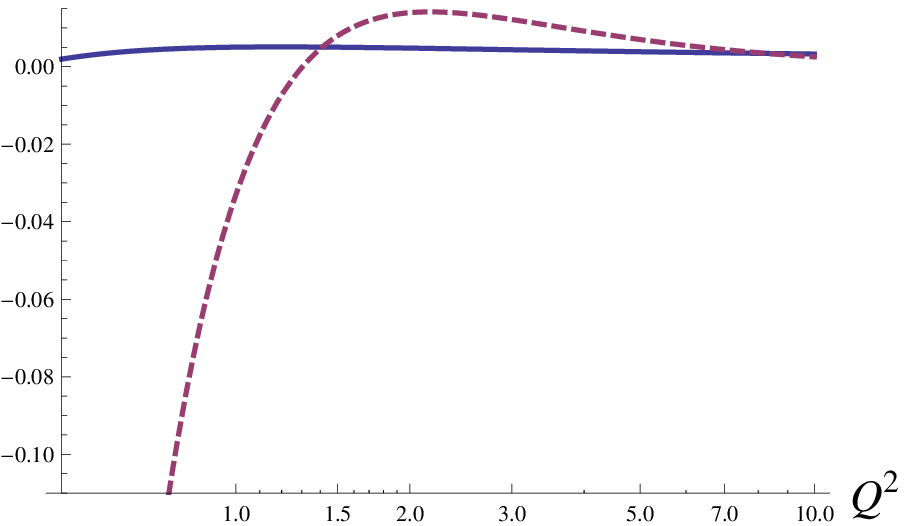}
 \caption{Relative difference between $F_2$ structure function and twists expansion for $x=0.001$.
 Left panel: $R_2^{(2)}$ contribution. Right panel: $R_2^{(2,4)}$ contribution. The solid lines follows from BK equation whereas
 the dashed curve from GBW model.
 Saturation scales from GBW model are $Q^2_{sat}=0.71$ GeV$^2$ for $x=0.001$.}
   \label{F2twist2_4}
 \end{center}
\end{figure}

It is also interesting to look at the relative content $R_2$ for $F_2$ structure function, defined as in (\ref{R_parameter}) with $\sigma_{T,L}$ replaced by $F_2$, (see Fig.\ \ref{F2twist2_4}). As was previously noticed \cite{twist1,twist2} higher twist contributions tend to cancel in this structure function. This pattern, driven by the twist structure of the $\gamma^*$ impact factor is found in both models.

All of the above results imply important consequences for experimental analysis of twist composition of proton structure. The BFKL/BK amplitude analysis suggests that the GBW model overestimates higher twists contribution to the total cross section. It would mean that an accurate determination of higher twist effects in DIS would require an enhanced experimental precision.

\section{Conclusions}

In this paper we analyzed the twist content of the proton structure functions in DIS within the framework of BFKL/BK equations and compared the obtained results with the GBW saturation model predictions \cite{twist1,twist2}. We confirmed that the higher twists contribution is more important for the longitudinal than for the transverse structure functions. We also found that for BFKL/BK the $F_2$ structure function is less susceptible for higher twist effects because of partial cancelation of twist-4 contribution (which is negative for $F_L$ and positive for $F_T$), as it was earlier found for the GBW model. There is, however, an important difference between the models in magnitude of higher twist corrections. The total cross-section in the BFKL/BK approach is strongly dominated by the leading twist contribution. In particular, the higher twists are strongly suppressed in the BFKL/BK calculation compared to the GBW predictions. This difference increases with decreasing $x$. The main reason is that at moderately low $Q^2$ the higher twists contribution decreases with decreasing $x$ in the BK framework and in increases with decreasing $x$ in the GBW model.
This prediction of the BK equation could provide a systematic explanation of why the higher twist effects in proton structure are relatively small and have not been found in DIS experiments yet. However, with enhanced sensitivity of the newest combined analysis of ZEUS and H1 DIS data including HERA2 results one sees some deviations from the leading twist description extrapolated towards small scales \cite{dis14talk} that might be a signature of higher twist effects and these deviations may be used to probe the models of higher twists in the proton structure.

There are two related issues that remain open. The first is a complete twist analysis of the BK amplitudes, that would have to treat not only the  $Q^2$-evolution in the double-logarithmic limit, as done in this paper, but also a more careful treatment of an impact of non-linear corrections on the input for the $Q^2$-evolution. Also important is to perform a twist decomposition of the diffractive DIS events withn BFKL/BK formalism. This should be quite interesting as recently an evidence of large higher twist contribution seems to emerge from ZEUS and H1 data \cite{ZEUSddis,MSS} and it calls for better understanding within QCD.

\textbf{Acknowledgement}

This research was supported by the Polish National Science Centre grant no. DEC-2011/01/B/ST2/03643.

\end{document}